\documentclass[twocolumn,aps,prb,superscriptaddress,floatfix]{revtex4}
\usepackage[T1]{fontenc}
\usepackage[utf8]{inputenc}
\usepackage[english]{babel}
\usepackage{color}
\usepackage{float}
\usepackage{braket}
\usepackage{amsmath}
\usepackage{amstext}
\usepackage{amssymb}
\usepackage{graphicx}
\usepackage{bbold}
\usepackage[unicode=true,pdfusetitle,
bookmarks=true,bookmarksnumbered=false,bookmarksopen=false,
breaklinks=false,pdfborder={0 0 1},backref=false,colorlinks=false]
{hyperref}
\hypersetup{
	colorlinks,linkcolor=blue,citecolor=blue,urlcolor=blue}
\usepackage{soul}

\makeatletter
\date{}

\makeatother
\setul{0.5ex}{0.3ex}
\definecolor{Blue}{rgb}{0,0.0,1}
\setulcolor{Blue}
\usepackage{babel}

\begin{document} 

\author{Tarik P. Cysne}
\email{tarik.cysne@gmail.com}
\affiliation{Instituto de F\'\i sica, Universidade Federal Fluminense, 24210-346 Niter\'oi RJ, Brazil} 

\author{Filipe S. M. Guimar\~{a}es}
\affiliation{J\"ulich Supercomputing Centre, Forschungszentrum J\"{u}lich and JARA, 52425 J\"{u}lich, Germany}

\author{Luis M. Canonico}
\affiliation{Catalan Institute of Nanoscience and Nanotechnology (ICN2), CSIC and BIST, Campus UAB, Bellaterra, 08193 Barcelona, Spain}

\author{Tatiana G. Rappoport}
\affiliation{Instituto de Telecomunicações, Instituto Superior Tecnico, University of Lisbon, Avenida Rovisco Pais 1, Lisboa, 1049001 Portugal}	
\affiliation{Instituto de F\'\i sica, Universidade Federal do Rio de Janeiro, C.P. 68528, 21941-972 Rio de Janeiro RJ, Brazil}

\author{R. B. Muniz}
\affiliation{Instituto de F\'\i sica, Universidade Federal Fluminense, 24210-346 Niter\'oi RJ, Brazil}
 
\title{Orbital magneto-electric effect in zigzag nanoribbons of $p$-band systems}

\begin{abstract}

Profiles of the spin and orbital angular momentum accumulations induced by a longitudinally applied electric field are explored in nanoribbons of $p$-band systems with a honeycomb lattice. We show that nanoribbons with zigzag borders can exhibit orbital magneto-electric effects. More specifically, we have found that purely orbital magnetization oriented perpendicularly to the ribbon may be induced in these systems by means of the external electric field, when sublattice symmetry is broken. The effect is rather general and may occur in other multi-orbital materials. 
\end{abstract}
\maketitle

\section{Introduction}

The magneto-electric effect (MEE) observed in certain materials evinces the interrelationship between their magnetic and electrical properties. 
It consists in either the appearance of a magnetization induced by an applied electric field or the advent of electric polarization brought about by an external magnetic field. In the literature distinct names have been given to the ME effect in order to highlight the main mechanisms involved and the relevant features of the systems where it manifests, but here we shall generically refer to them simply as magneto-electric effect (MEE), without detriment to their specificities.~\cite{kinetic_ME-Insulators_1,kinetic_ME-metals_1, kinetic_ME-metals_2, ME-Spin-Orbital-contrib}.

Possibilities of utilizing electric field to control the magnetization (or the other way around) are certainly of great interest for device applications~\cite{Fusil:2014,Ortega:2015,Cheng:2018,Hu:2019,adiabaticEvolution}. The MEE has been explored in various materials, including antiferromagnetic systems~\cite{ME-antiferromagnets-1, ME-antiferromagnets-2}, multiferroic composites~\cite{ME-multiferroic-1, ME-multiferroic-2}, and topological insulators~\cite{ME-topologicalInsulators, ME-topologicalInsulators_2, ME-topologicalInsulators_3}, among others. From a microscopic point of view, contributions to the MEE may originate from the electronic spin and/or from its orbital angular momentum (OAM)~\cite{ME-Spin-Orbital-contrib}. In the second case, the phenomenon is often called the orbital magneto-electric effect (OME)~\cite{Levitov-1985}, which has also been previously investigated, for example, in helical lattices~\cite{Orbital-Edelstein, OrbEdelstein_2, Murakami_OME_woutInversionSymmetry}, graphene twisted bilayers~\cite{OME_tBG-2020} and magnetic nanoparticles~\cite{ME-Nanoparticles}.

An external electric field may also induce the appearance of an OAM current flowing transversely to the applied field direction. This is the so-called orbital-Hall effect (OHE), which is similar to the spin-Hall effect (SHE)~\cite{Sinova_RA} but, contrarily to the latter, does not necessarily requires the presence of spin-orbit interaction to occur. Some years ago, the OHE was predicted to happen in semiconductors~\cite{OHEBernevig} and in metals, where it can be very strong~\cite{OHE_Metals_0, OHE_Metals_2, OHE_Metals_1}. More recently, significant interest in the OHE and other orbital phenomena have been revived~\cite{Go_2020PRL,Orbital-torque,GoXueLinearResponse, xiao2020detection,OrbitalTexture, Bhowal_OME, OHEmetals_Texture, beaulieu2020revealing, schuler2021bloch}, specially in two-dimensional (2D) systems~\cite{OrbitalPhotocurrents, GraphaneOHE, OrbitalTextureBorophene, Mele-OrbitalHall, Orbital_Rashba_2, Orbital_Rashba_1, LuisPRL, Uspxpy, UsTmd, BhowalTMD, bhowal2021orbital, Xue-Imaging_VHE_OHE, Us_TMD-bilayer}, with the aim of exploring novel possibilities for utilizing OAM currents to transmit and store information on nanoscopic scales \cite{OrbitalPhotocurrents, xiao2020detection}. 

The spin and orbital Hall effects may lead to angular momentum accumulations at the system's boundaries. 
In topological insulators, these borders host conducting electronic states that are topologically protected by symmetry and thus robust to inhomogeneities.
Nanoribbons of these materials exhibit these edge states and are quite useful for investigating their features. Recently, we have examined the transport properties of charge, spin and OAM in $p_x$-$p_y$ orbital systems with a 2D honeycomb lattice~\cite{LuisPRL, Uspxpy}. This model was introduced and intensively explored in the context of optical lattices, where it is possible to filter the $p_z$-orbital by application of laser beams polarized in the z-direction, leaving only the $p_x$ and $p_y$ orbitals effectively active \cite{Wu-pxpy-2008, OpticalLattices_1, OpticalLattices_2, OpticalLattices_3, OpticalLattices_4, OpticalLattices_5, OpticalLattices_6, PhotonicRibbonModel}. More recently, it has been found that this relatively simple model also describes fairly well the low energy electronic properties of group-V-based 2D materials grown atop a SiC substrate. Here, the $p_z$-orbital filtering occurs naturally due to the interaction between the overlayer and the SiC substrate~\cite{groupV-Li-PRB-2018, pxpy-BiSi-MaterialRealization_1, pxpy-Bi-MaterialRealization_2, ReisBismutheneExperimental, pxpy-Antimonene-MaterialRealization_1}. Despite its simplicity, the $p_x$-$p_y$ tight-binding model on a honeycomb lattice exhibits a rich topological phase diagram as a function of the spin-orbit coupling and the sublattice asymmetry potential strengths~\cite{MultiorbitalHoneycomb-Wu}.
We have analysed some of these phases and under certain circumstances found sizeable OHE, with values that exceed those obtained for the SHE~\cite{Uspxpy}.

Quantum confinement effects caused by the nanoribbons finite width may significantly alter the electronic states and transport properties of a system~\cite{NR-gap1, NR-gap2, SLouie_NR_2006, PhysRevResearch.3.033021}. 
Quite generally, they depend upon the ribbon's breadths and edge shapes, enabling the emergence of novel attributes and functionalities~\cite{Halfmetalic_NR, FDominguez-TestingTOpologicalprotection, Guinea_TMD_NR}.  
Hence, it is instructive to inquire into how nanoribbons of these $p_x$-$p_y$-orbital materials respond to an electric field applied along the stripe axis and, in particular, evaluate the spin and OAM disturbances induced by it. 

Here we calculate profiles of the spin and OAM accumulations produced by a longitudinally applied electric field on nanribbons of $p$-band systems. We show that a MEE can take place in zigzag (ZZ) nanoribbons of these materials where a purely orbital magnetization is brought about by this external field. 
The possibility of controlling the appearance of orbital magnetization by means of an electric field enlarge the prospects of using these systems for orbitronic applications.   
 
\section{Model and methods}

We consider a tight-binding model with two atomic orbitals ($p_{x}, p_{y}$) per site on a honeycomb lattice~\cite{MultiorbitalHoneycomb-Wu, groupV-Li-PRB-2018}, described by the Hamiltonian
\begin{eqnarray}
\mathcal{H}&=&\sum_{\langle ij\rangle}\sum_{\mu \nu s} \left( t^{\mu\nu}_{ij} +\epsilon_i\delta_{ij} \delta_{\mu\nu} \right) p^{\dagger}_{i \mu s}p_{j \nu s} \nonumber \\
&& +\lambda_{I} \sum_{i \mu \nu  s} s_{ss}^z L_{\mu,\nu}^zp^{\dagger}_{i \mu s}p_{i\nu s},
\label{Hpxpy}
\end{eqnarray}
where the first line represents the electronic kinetic energy plus a spin independent local potential, and the second one describes the intrinsic atomic spin-orbit interaction in the subspace spanned by the $p_x$-$p_y$ orbitals.   
Here, $i$ and $j$ denote the honeycomb lattice sites positioned at $\vec{R}_i$ and $\vec{R}_j$, respectively. 
The symbol $\langle ij \rangle$ indicates that the sum runs over nearest-neighbour atoms only. 
The operator $p^{\dagger}_{i \mu s}$ creates an electron of spin $s=\uparrow,\downarrow$ in the atomic orbitals $p_{\mu}\, (\mu=x,y)$ located at $\vec{R}_i$. 
$\epsilon_i$ represents the on-site atomic energy plus a staggered local potential $V_i$ that breaks the inversion symmetry between the two interpenetrating triangular sublattices A and B; $V_i = + V_\text{AB}$ ($- V_\text{AB}$), when $i$ belongs to sublattice A (B).
The hopping integrals $t_{ij}^{\mu\nu}$ between orbitals $p_{i \mu}$ and $p_{j \nu}$ are parametrized according to the standard Slater-Koster tight-binding formalism~\cite{Slater-Koster} and may be expressed in terms of the usual two-center integrals $V_{pp\sigma}$ and $V_{pp\pi}$. 
$\lambda_I$ represents the strength of the intrinsic SOC, $s^z_{ss}$ are the diagonal elements of the usual Pauli matrix $\text{diag}(1,-1)$, and $L_{\mu,\nu}^z$ are matrix elements of OAM operator represented in $p_{x,y}$ subspace. More details on the Hamiltonian can be found in Appendix~\ref{appA}.

\begin{figure}[h]
	\centering
	 \includegraphics[width=0.9\linewidth,clip]{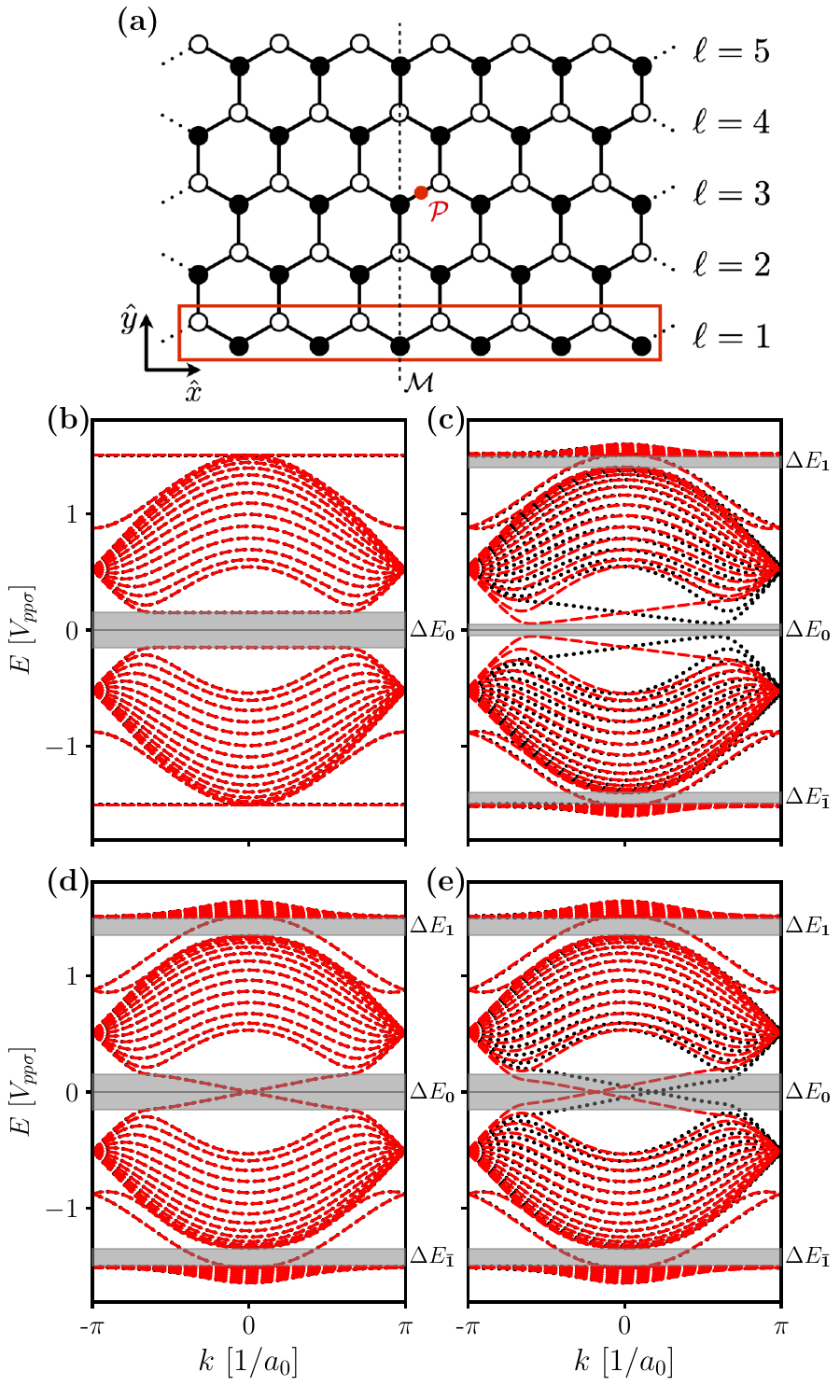}   
  
	\caption{(a) Schematic representation of a ZZ nanoribbon comprising 5 atomic lines that are identified by integers numbers $\ell = 1, .., 5$; the atoms in sublattice A (B) are depicted in white and black circles, respectively. The vertical dashed black line indicates the plane of mirror symmetry ($\mathcal{M}$) which contains an AB-site vertical bond in ZZ nanoribbon. The red circle indicates the spatial-inversion symmetry center ($\mathcal{P}$) of the ZZ nanoribbon in the absence of sublattice potential. Panels (b) to (e) illustrate the calculated electronic energy bands for a ZZ nanoribbon with 15 lines in breadth. The wave-vector $k$ is written in units of $a_0^{-1}$, where $a_0$ represents the lattice constant. Black dotted and red dashed lines refer to $\uparrow$- and $\downarrow$-spin bands, respectively. Results calculated for $V_\text{AB}=0.15$ with $\lambda_I=0$ and $0.10$ are displayed in panels (b) and (c), respectively. Panel (d) shows the results calculated with $\lambda_I=0.15$ and $V_\text{AB}=0$, while panel (e) displays the results obtained with $\lambda_I=0.15$ and $V_\text{AB}=0.05$. The gray stripes highlight the 2D bulk energy-band gaps denoted by $\Delta E_{\bar{1}}$, $\Delta E_0$, and $\Delta E_1$.}
	\label{fig:figA2}
\end{figure} 
  
In our calculations, the energy origin is chosen to coincide with the energy level of the atomic orbitals $p_{x,y}$.  For simplicity, we assume that $V_{pp\pi}=0$ and take $V_{pp\sigma}=1$ as our energy unit. 
We shall explore the accumulations of spin and OAM induced by an electric field applied along the nanoribbon axis direction. Fig.~\ref{fig:figA2}(a) illustrates a nanoribbon with ZZ edges in which the atoms belonging to sublattices A and B are represented by white and black circles, respectively. It is finite along the transverse $\hat{y}$ direction and, in general, have $N$ atomic lines in breadth, which are identified by integer numbers $\ell=1,..,N$, as schematically illustrated in Fig.~\ref{fig:figA2}(a) for $N=5$.
Panels (b) to (e) of Fig.~\ref{fig:figA2} show the calculated band-energy spectra for ZZ nanoribbons with 15 atomic lines in breadth for different values of $V_\text{AB}$ and $\lambda_I$. 
Panels (b) and (c) display results for nanoribbons taken from 2D bulk systems in which the central energy 
band-gap $\Delta E_0$ is non-topological. This is confirmed by the absence of edge states crossing this energy range, where the system is an ordinary insulator. For $\lambda_I=0.10$ and $V_{\text{AB}}=0.15$ depicted in panel (c), the bulk system is in the A1 phase, which is categorized by the set of $\uparrow$-spin band Chern numbers $\mathcal{C}^{\text{A1}}_s= (1, -1, 1, -1)$ in Ref.~\onlinecite{MultiorbitalHoneycomb-Wu}. In this case, the system exhibits two lateral energy band gaps ($\Delta E_{\bar{1}}$ and $\Delta E_1$) that are topological, in addition to the non-topological one $\Delta E_0$. Panel (c) show that $\Delta E_{\bar{1}}$ and $\Delta E_1$ are crossed by chiral edge states in the ribbon geometry, as expected~\cite{Kane-Mele-2005-1}. Panels (d) and (e) of Fig.~\ref{fig:figA2} display the energy bands for nanoribbons extracted from 2D systems in the B1 phase, classified by $\mathcal{C}^{\text{B1}}_s= (1, 0, 0, -1)$, where they exhibit three topologically nontrivial bulk energy-band gaps, within which the systems behave as a quantum spin Hall insulator~\cite{MultiorbitalHoneycomb-Wu, Uspxpy}. We clearly see in panels (d) and (e) that all three gaps are crossed by edge states. As depicted in panel (b) and expected from our non-interacting particles approach, the energy-band spectrum for $\lambda_I=0$ is spin degenerated. The same happens in the absence of $V_\text{AB}$, as exemplified in panel (d). This is due to a combination of time-reversal symmetry and spatial-inversion. However, when sublattice symmetry is broken, SOC lifts this degeneracy giving rise to spin-split energy bands, as illustrated in panels (c) and (e). 

We shall calculate the spin $\delta \langle \hat{S}^z_\ell \rangle$ and OAM $\delta \langle \hat{L}^z_\ell \rangle$ accumulations per atom induced at each atomic line $\ell$ in the nanoribbons by an applied electric field. For this purpose, we utilize linear response theory following the same procedure described in Refs.~\citenum{Orbital-torque} and \citenum{GoXueLinearResponse}.
The electric field is applied in-plane, along the longitudinal $\hat{x}$ direction, as depicted in Fig.~\ref{fig:figA2}(a).
Up to linear order in the perturbing field, the inter-band and intra-band contributions to the change in the expectation value of an observable $\delta \langle\hat{\mathcal{O}}_{\ell}\rangle$ due to the applied field are given by,
\begin{subequations}
\begin{equation}
\begin{split}
\delta \langle\hat{\mathcal{O}}_{\ell}\rangle^\text{Intra}=&-\frac{e\hbar \mathcal{E}_x}{2\Gamma} \sum_{n,k}\frac{\partial f_{nk}}{\partial E} \\
& \times \big<nk\big|\hat{\mathcal{O}}_{\ell}\big|nk\big>\big<nk\big|v(k)\big|nk\big>, \label{Ointra}
\end{split}
\end{equation}
\begin{equation}
\begin{split}
\delta \langle\hat{\mathcal{O}}_{\ell}\rangle^\text{Inter}=&e\hbar \mathcal{E}_x \sum_{n,m,k} (f_{nk}-f_{mk})\\
& \times \text{Im}\Bigg[\frac{\big<nk\big|\hat{\mathcal{O}}_{\ell}\big|mk\big>\big<mk\big|v(k)\big|nk\big>}{(E_{nk}-E_{mk}+i\eta)^2}\Bigg].
\label{Ointer}
\end{split}
\end{equation}
\end{subequations}
Here, $\hat{\mathcal{O}}_{\ell}$ may represent either the spin $\hat{S}_{\ell}^z$ or the OAM $\hat{L}_{\ell}^z$ operators projected on line $\ell$~\cite{GManchonAManchon-2020}. For the OAM operator, we follow Ref. \onlinecite{GoXueLinearResponse} and use the intra-atomic orbital approximation. $E_{nk}$ are the eigenvalues and $\ket{nk}$ the corresponding eigenvectors of the Hamiltonian given by Eq.~\eqref{Hpxpy} evaluated in the reciprocal space; $n$ denotes the energy band index, $k$ is the wave vector, and $f_{nk}$ symbolizes the Fermi-Dirac distribution function associated with the state $\ket{nk}$. $\hat{v}(k)=\hbar^{-1}(\partial \hat{H}(k)/\partial k)$ is the velocity operator, $e$ is the modulus of the electronic charge, and $\mathcal{E}_x$ denotes the intensity of the applied electric field. $\Gamma = \hbar/(2\tau)$, where $\tau$ is the momentum relaxation time, is treated here as a phenomenological parameter that simulates effects of disorder in the transport properties of the nanoribbons within the constant relaxation-time approximation~\cite{GoXueLinearResponse, Kubo-Manchon}; $\eta$ is a small positive quantity arising from a conventional artifice to ensure that the external perturbation is turned on adiabatically.

\section{Results and Discussions}

\begin{figure}[h!]
	\includegraphics[width=1.00\linewidth]{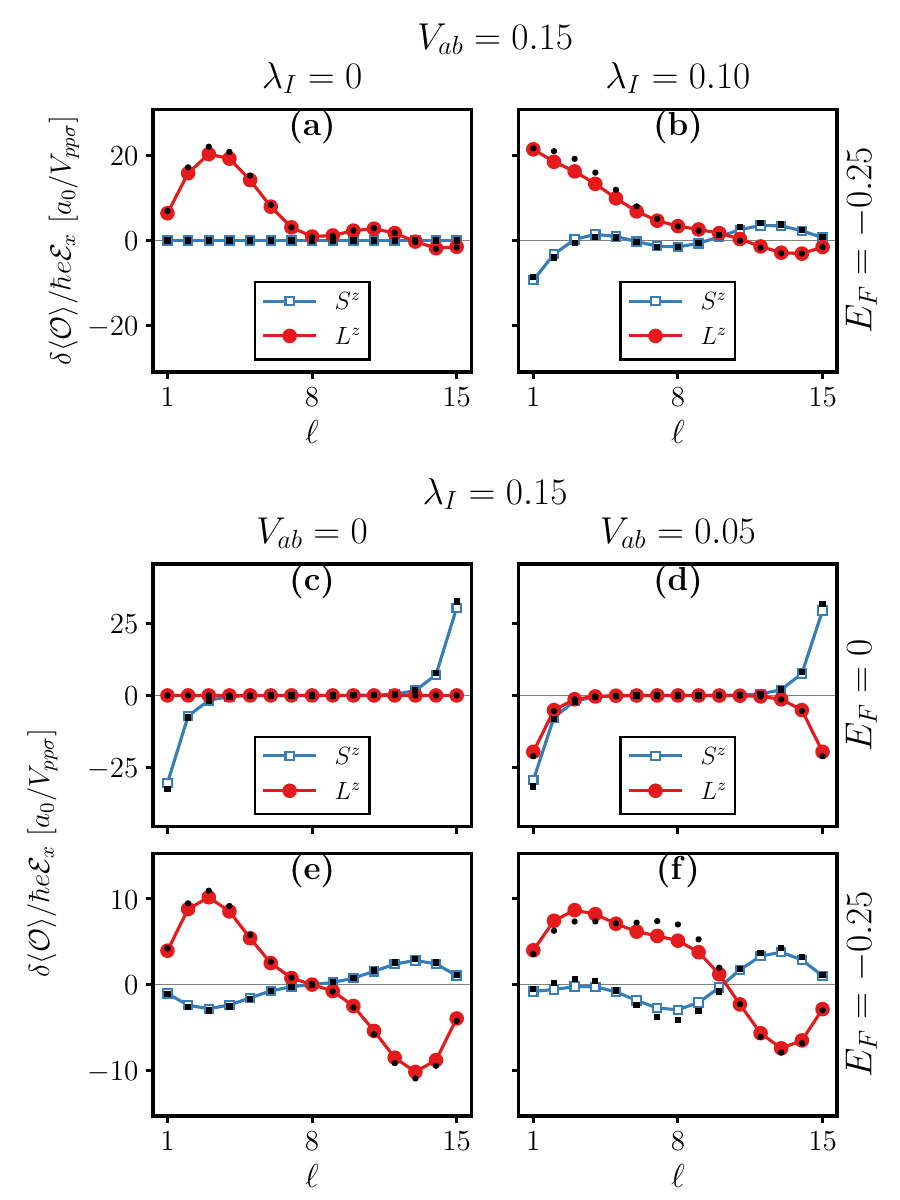}
	\caption{ Spin and OAM accumulation profiles induced by an applied electric field across ZZ nanoribbons with 15 atomic lines in breadth. $\delta \langle S_\ell^z\rangle$ and $\delta \langle L_\ell^z\rangle$ are calculated per atom in each atomic line $\ell$ of the ribbon. 
	 Panels (a) and (b) illustrate results for $V_\text{AB}=0.15$ and $E_F=-0.25$, with $\lambda_I=0$ and $0.10$, respectively. Results obtained for $V_\text{AB}=0$, and $\lambda_I=0.15$, with $E_F = 0$, and $E_F = -0.25$ are depicted in panels (c) and (e), respectively. 
	Panels (d) and (f) display results for $V_\text{AB}=0.05$, $\lambda_I=0.15$ and the same values of $E_F$. Black squares and circles represent the results obtained by means of an alternative approach (see text) for the spin and OAM accumulations, respectively. Here, we used $\Gamma=10^{-3}$.}
	\label{fig:fig_zznrResult1} 
\end{figure}

We consider a ZZ nanoribbon with 15 atomic lines and calculate the spin and OAM accumulations per atom induced by the applied electric field in each one of these lines. We start with the two cases in which the central energy-band gap $\Delta E_0$ of the 2D system is non-topological, namely when $V_\text{AB}=0.15$ with $\lambda_I=0$ and $0.10$, whose energy bands are displayed in Figs.~\ref{fig:figA2}(b) and (c), respectively. In these circumstances, both the induced spin and OAM accumulations vanish for $E_F=0$, which is compatible with the fact that the bulk 2D system for those sets of parameters is an ordinary insulator that shows neither SHE nor OHE at this Fermi energy (see Ref.~\onlinecite{Uspxpy}). However, beyond the energy gap $\Delta E_0$ (e.g. for $E_F=-0.25$), the OAM accumulation is not zero, as panels (a) and (b) of Fig.~\ref{fig:fig_zznrResult1} illustrate. Panel (a) reveals that outside the range of $\Delta E_0$ we obtain OAM accumulation throughout the ribbon, even in the absence of SOC. While the induced spin accumulation in this case vanishes (as expected), the OAM profile is finite. Its most striking feature is the lack of symmetry with respect to the ribbon axis, which is markedly different from what one would expect solely from the OHE. Moreover, $\delta \langle L^z_t\rangle=\sum_{\ell} \delta \langle L^z_\ell\rangle\neq 0$, indicating the appearance of an induced net orbital magnetization---typical of an orbital magneto-electric effect. Generally, the magnetization originates from spin and orbital magnetic moments, but in this specific case there is no spin contribution and the induced magnetization has solely orbital character. Panel (b) shows that, for a nanoribbon extracted from a 2D system in the A1 phase, both the induced spin and OAM accumulation profiles are finite in the metallic regime and also \emph{asymmetric} with respect to the ribbon axis. By summing the contributions of all lines, we obtain that the net magnetization remains purely orbital, as the total spin contribution vanishes.

We shall now explore some cases in which the nanoribbons are extracted from 2D systems in the B1 phase, where all three energy-band gaps are topological, as illustrated in Figs.~\ref{fig:figA2}(d) and (e). For $\lambda_I=0.15$ and in the absence of $V_\text{AB}$, we see in panel (c) of Fig. ~\ref{fig:fig_zznrResult1} that sizeable spin accumulations with opposite signs appear near the nanoribbon's edges for $E_F=0$, whereas $\delta \langle L^z_\ell\rangle = 0 ~\forall~\ell$. This is what one would expect from the SHE and is also consistent with the fact that in the B1 phase the 2D system behaves as a quantum spin Hall insulator that displays no orbital Hall effect within $\Delta E_0$~\cite{Uspxpy}. In contrast, by moving the Fermi level outside of $\Delta E_0$ to $E_F=-0.25$, we observe in Fig.~\ref{fig:fig_zznrResult1}(e) that the induced OAM is finite. Here, both the induced spin and OAM accumulations are much more spread across the ribbon and not so restricted to the borders as in the case of $E_F=0$. Nevertheless, they are both \emph{anti-symmetric} with respect to the ZZ ribbon axis and hence do not lead to a net magnetization, as the spin and orbital Hall effects forecast~\cite{Uspxpy}. Moreover, when the sublattice symmetry is broken by a staggered potential $V_{\text{AB}}=0.05$, different profiles emerge. For example, in Fig.~\ref{fig:fig_zznrResult1}(d), we obtain an induced spin accumulation profile for $E_F=0$ that is rather similar to the previous case, having anti-symmetric character with respect to the nanoribbon axis as one would expect from the SHE. However, the induced OAM profile across the ribbon is \emph{symmetric} with respect to the nanoribbon axis, which is not expected from the OHE and clearly leads to a non-zero net orbital magnetization when we sum over all lines. In panel (f) of Fig.~\ref{fig:fig_zznrResult1}, asymmetric profiles for both the induced spin and OAM accumulations are also revealed for $E_F=-0.25$. Here, by summing over all lines the spin contribution once again vanishes and the resulting induced magnetization acquires a purely orbital character.

In order to confirm our predictions, we have repeated our calculations employing an alternate approach, described in Refs.~\citenum{PhysRevB.92.220410} and~\citenum{Guimaraes2017}. It considers a spatially uniform and time dependent harmonic electric field with small amplitude $\mathcal{E}_x$ that within linear response theory gives rise to a local spin and OAM disturbances per atom in line $\ell$ given by 
\begin{equation}
\delta\langle \mathcal{O}_{\ell}(t)\rangle =-\lim_{\omega\rightarrow0}\frac{e\mathcal{E}_x}{\hbar\omega}\operatorname{Im}\left\{e^{-i\omega t}\mathcal{D}_{\ell}(\omega)\right\}\ , \label{StaticSucep1}
\end{equation}
 where
\begin{equation}
\mathcal{D}_{\ell}(\omega) = \sum_{\substack{k\\s_1s_2s_3}}\sum_{\substack{\ell_1\ell_2\\\mu\nu\gamma\xi}}\mathcal{O}^{s_1 s_2}_{\mu \nu} \chi^{^{s_1 s_2 s_3 s_3}\mu\nu\gamma\xi}
_{\ell\ell\ell_1\ell_2}(k,\omega)~\frac{\partial t_{\ell_2\ell_1}^{\gamma\xi}(k)}{\partial k}\ . \label{StaticSucep2}
\end{equation}
Here, $s_1$, $s_2$ and $s_3$ designate the $\uparrow$ and $\downarrow$ spin directions, $\mu$, $\nu$, $\gamma$, and $\xi$ denote the $p_x$ and $p_y$ atomic orbitals, and $\ell$, $\ell_1$, and $\ell_2$ label the atomic lines. The matrix elements $\mathcal{O}^{s_1 s_2}_{\mu \nu} = S^z_{s_1 s_2} \delta_{\mu\nu}$ for the spin accumulation or $\mathcal{O}^{s_1 s_2}_{\mu \nu} = L_{\mu\nu}^{z}\delta_{s_1 s_2}$ for the OAM disturbance. In our case, since we are neglecting electronic Coulomb interaction, $\chi(k,\omega)$ represent generalized non-interacting spin susceptibilities. The results obtained with this method for the induced spin and OAM accumulations are also depicted in Fig.~\ref{fig:fig_zznrResult1} by black squares and circles, respectively. The agreement between the two approaches is excellent and corroborates our findings. 
  
\begin{figure}[h!]    
	\includegraphics[width=1.00\linewidth]{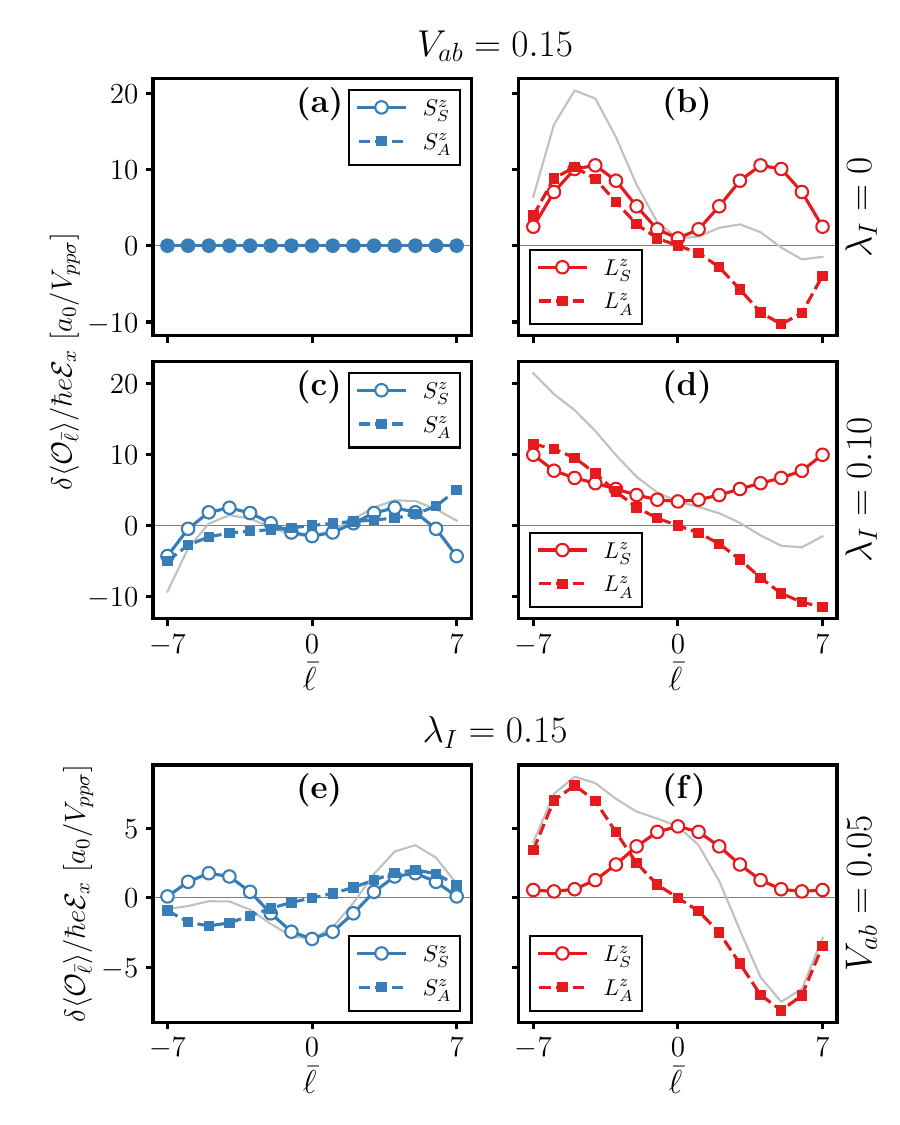}
	\caption{Decomposition into symmetric and anti-symmetric components of the induced spin and OAM profiles depicted in panels (a), (b) and (f) of Fig. \ref{fig:fig_zznrResult1}. The original profiles are reproduced here by the grey lines.
	Panels (a) and (b) show the decomposition of the spin and OAM profiles depicted in panel (a) of Fig. \ref{fig:fig_zznrResult1}, calculated for $V_\text{AB}=0.15$, $\lambda_I=0$, and $E_F=-0.25$.
	Panels (c) and (d) show the decomposition of the spin and OAM profiles depicted in panel (b) of Fig. \ref{fig:fig_zznrResult1}, calculated for $V_\text{AB}=0.15$, $\lambda_I=0.10$, and  $E_F=-0.25$.
	Panels (e) and (f) show the decomposition of the spin and OAM profiles depicted in panel (f) of Fig. \ref{fig:fig_zznrResult1}, calculated for $V_\text{AB}=0.05$, $\lambda_I=0.15$ and $E_F=-0.25$.
	}
	\label{fig:fig_zznr_AS} 
\end{figure}         

To extract some physical insights into the different effects that contribute to the spin and OAM responses on ZZ nanoribbon, we further explore the cases with asymmetric profiles exhibited in panels (a), (b) and (f) of Fig.~\ref{fig:fig_zznrResult1}. It is instructive to identify their symmetric and anti-symmetric components. To this end it is useful to relabel the atomic lines by $\bar{\ell} \in [-7, 7]$. The anti-symmetric $(-)$ and symmetric $(+)$ contributions to the profiles are obtained by $\delta\langle\hat{\mathcal{O}}_{\bar{\ell}}\rangle^{\mp}=\big(\delta\langle\hat{\mathcal{O}}_{\bar{\ell}}\rangle\mp \delta\langle\hat{\mathcal{O}}_{-\bar{\ell}}\rangle\big)/2$, and are shown in Fig.~\ref{fig:fig_zznr_AS}. We identify the anti-symmetric contributions to the induced spin ($\langle \delta S^z\rangle_A$) and OAM ($\langle \delta L^z\rangle_A$) profiles as arising from the spin and orbital Hall effects \cite{Uspxpy}, respectively, and the symmetric ones ($\langle \delta S^z\rangle_S$ and $\langle \delta L^z\rangle_S$) are ascribed to the MEE. Clearly, a nonzero induced magnetization can only come from the symmetric components. However, the appearance of a symmetric contribution does not necessarily lead to a nonzero magnetization. For example, the symmetric spin contributions to the total induced magnetization illustrated in panels (c) and (e) of Fig.~\ref{fig:fig_zznr_AS} is zero, i.e., $\sum_{\bar{\ell}=-7}^7 \langle \delta S_\ell^z\rangle_S=0$. In contrast, the symmetric orbital components depicted in panel (b), (d) and (f) of the same figure clearly lead to a non-zero magnetization that is purely orbital, characterizing an orbital magneto-electric effect.

\begin{figure}[h!]    
	\includegraphics[width=1.00\linewidth]{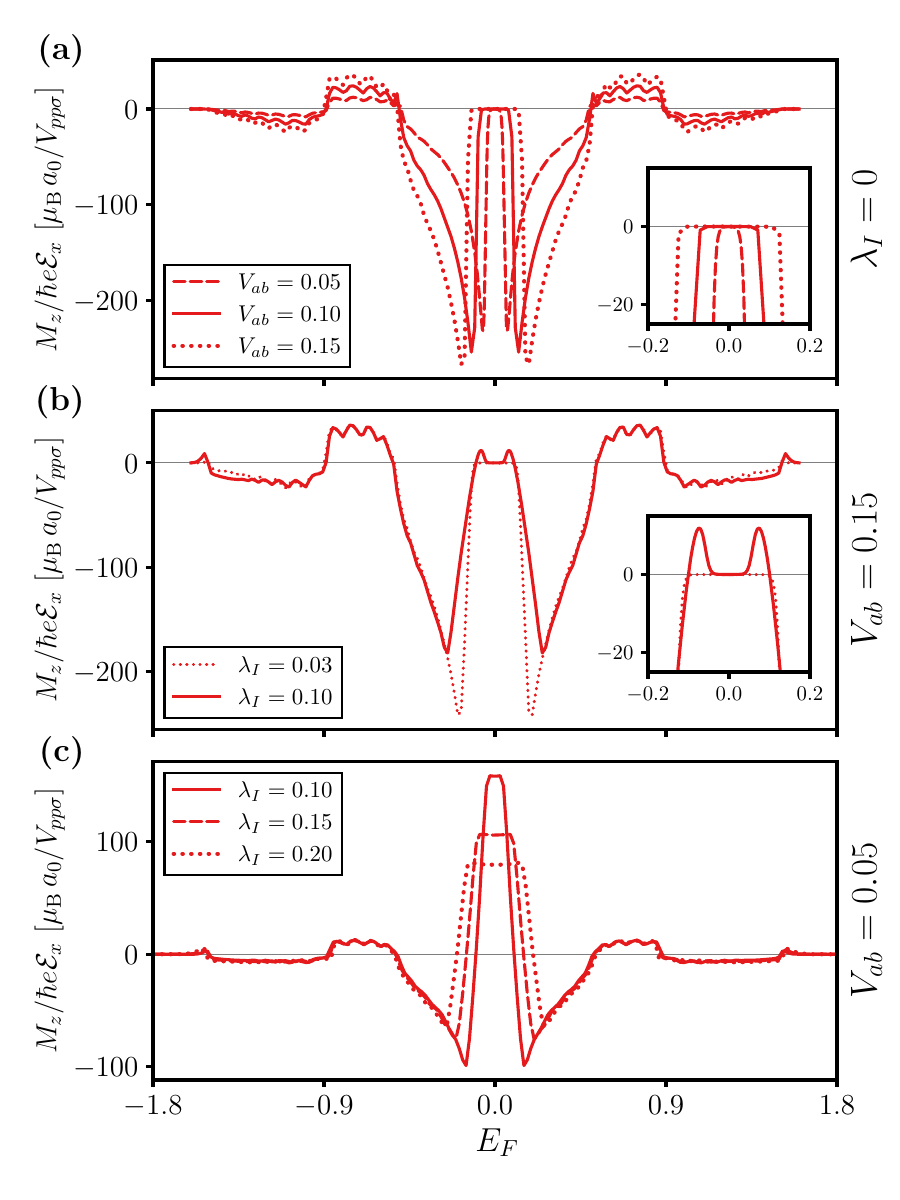}
	\caption{Electric-field induced orbital magnetization ($M_z$) per unit cell calculated as a function of $E_F$ for a ZZ nanoribbon with 15 lines. Panel (a) show results obtained for $\lambda_{I}=0$ and three distinct values of $V_{\text{AB}}$. The inset highlights the calculated values of $M_z (E_F)$ around $E_F=0$. In panel (b) we display results for nanoribbons taken from a 2D system in the A1 phase, calculated for $V_{\text{AB}}=0.15$ with $\lambda_I = 0.03$ and $0.10$. Panel (c) exhibits results calculated for nanoribbons taken from a 2D system in the B1 phase with $V_{AB}=0.05$ for three different values of $\lambda_{I}$.
	\label{fig:fig4}} 
\end{figure}

We shall now examine how the total induced orbital magnetization per unit cell of the ZZ nanoribon ($M_z = -\mu_{\text{B}} \langle \delta L^z_t\rangle$) is influenced by the sublattice symmetry breaking potential $V_{\text{AB}}$, the SOC strength $\lambda_I$ and the Fermi energy $E_F$. Here $\mu_{\text{B}}$ denotes the Bohr magneton. We start with the case in which $\lambda_I = 0$, where $\Delta E_0$ is non-topological and is not crossed by conducting edge states in the ribbon geometry. Fig.~\ref{fig:fig4}(a) shows results of $M_z$ calculated as a function of energy for different values of $V_{\text{AB}}$. The energy-band gaps increase with $V_{\text{AB}}$, and within them the OME vanishes. However, outside and close to the energy-gap borders, the OME assumes quite large values even in the absence of SOC, indicating that with the use of a gate voltage one may control the advent of the OME in this case. In panel (b) of Fig.~\ref{fig:fig4}, we show results calculated for $V_{\text{AB}}=0.15$ and $\lambda_I = 0.03$ and $0.10$, which represent nanoribbons extracted from a 2D system in the A1 phase. Here we see once more that for energies within the corresponding central energy-band gaps, where the system is an ordinary insulator, there is also no OME. Nevertheless, in the metallic regime, the OME may reach relatively high values that increase as the SOC diminishes. Finally, in panel (c) we display results calculated for $V_{\text{AB}}=0.05$ and three different values of $\lambda_I$. They all refer to nanoribbons extracted from a 2D system in the B1 phase where, in contrast with the two previous cases, the central energy-band gap is topological and crossed by conducting edge states in the stripe geometry, as exhibited in panels (d) and (e) of Fig.~\ref{fig:figA2}. We clearly see that the OME is finite and relatively large within $\Delta E_0$, changing sign when $E_F$ crosses its borders, and also attaining fairly high values outside the range of $\Delta E_0$. Once again, we note that in this phase, which requires $\lambda_I > V_{\text{AB}}$, the maximum intensity of the OME also reduces as the SOC increases. Here, it seems also possible to manipulate the direction of the induced orbital magnetization with the of use of gate voltages.

It is noteworthy that the appearance of the orbital magneto-electric effect in these nanoribbons requires sublattice symmetry breaking. Spin-orbit coupling influences the OME but is not a necessary ingredient for its occurrence. Our results reveal that with the increase of the SOC strength the maximum intensity of the OME actually reduces. They also show that the OME vanishes in the absence of conducting states at the Fermi energy, which typifies a current-induced magnetization effect, also called kinetic magneto-electric effect \cite{kinetic_ME-metals_1, kinetic_ME-metals_2}. This is consistent with the fact that the induced orbital magnetization is totally dominated by the intra-band contribution [Eq.~\eqref{Ointra}] for the cases we have examined. Up to first order in the applied field, the magneto-electric effect is well described by the magneto-electric susceptibility tensor $\hat{\alpha}$, which links the components of the induced magnetization to the applied electric field: $M_i = \sum_j \alpha_{ij} \mathcal{E}_j$, where $i$ and $j$ here denote the Cartesian directions $x,y,z$, and $\alpha_{ij}$ represent the matrix elements of $\hat{\alpha}$. 
The general form of $\hat{\alpha}$ may be determined by symmetry arguments \cite{Hayami_2014_OME}. Parity and time reversal symmetries in particular play an important role in the magneto-electric effect. For example: it is possible to show that the intra-band contribution to $M_z$ requires inversion symmetry to be broken, as pointed out in Refs.~\onlinecite{TemperatureGradient} and~\onlinecite{Satoru-KuboTR}. However, to activate the inter-band contribution [Eq.~\eqref{Ointer}] it is also necessary to break time-reversal symmetry. This justifies the absence of inter-band contributions in our calculations, since our Hamiltonian breaks spatial inversion symmetry in the presence of $V_{\text{AB}}$, but remains invariant by time-reversal. It is also worth mentioning that the crystal structure of the 2D bulk system in the presence of $V_{\text{AB}}$ belongs to the point group $D_{3h}$ that leads to $\alpha_{zx}=0$~\cite{PhysRevResearch.3.023111} and hence does not allow the OME to take place. Nevertheless, for the zigzag nanoribbon with sublattice asymmetry, the point group is reduced to $C_{2v}$, with the principal axis lying in-plane along the $\hat{y}$ direction, which allows non-zero values of $\alpha_{zx}$~\cite{PhysRevResearch.2.012073}. 

In Fig.~\ref{fig:fig4} (a) and (b) we see that the maximum calculated value for the total induced orbital magnetization per unit cell of the ZZ nanoribon $M_z/(\hbar e\mathcal{E}_x)$ is approximately $200 \mu_{\text{B}} a_0/V_{pp\sigma}$ in the metallic regime close to the energy-band-gap borders. Thus, for $a_0/V_{pp\sigma} \approx 0.25$nm/eV (typical of the group-V elements/SiC) and for $\mathcal{E}_x = 1\times10^5$V/m, we obtain a value of $M_z \approx 5\times10^{-3}\mu_{\text{B}}$. This is more than one order of magnitude larger than the current-induced magnetic moment for Au(111), and is slightly larger than the results obtained for Bi/Ag(111) and for the $\alpha$-Sn(001) surface~\cite{PhysRevB.97.085417}. The results depicted in Fig.~\ref{fig:fig4}(c) show that the absolute value of $M_z$ for energies within the bulk topological energy-band gap $\Delta E_0$ is a little bit smaller but still comparable with the previous case for the same value of $\Gamma = 10^{-3}$, which corresponds to a momentum relaxation time $\tau \approx 0.17 \text{ps}$. In this case, it is worth noting that the appearance of the induced magnetization is mediated by conducting edge states that are topologically protected against disordered scattering. Therefore, the induced magnetization may be much larger, since it increases linearly with $\tau$, according to Eq.~\eqref{Ointra}.

Here, since the OME is mediated by conducting states only, it may be useful to associate the induced magnetization with the components of the electric-current density $\mathcal{J}_i$ as $M_i = \sum_j \beta_{ij} \mathcal{J}_j$ \cite{Orbital-Edelstein}. In our case $M_z = \beta_{zx} \mathcal{J}_x$, where $\beta_{zx} = \alpha_{zx} \rho_{xx}$ with $\rho_{xx}$ representing the longitudinal resistivity. Since $\alpha_{zx} \propto \tau$ and 
$\rho_{xx} \propto 1/\tau$, $\beta_{zx}$ does not depend upon $\tau$. 
It is also noteworthy that the staggered local potential ($V_i = \pm V_{AB}$) creates electric dipoles that cancel out in the bulk but not for nanoribbons with zigzag edges, where a net in-plane polarization $P_y$ emerges along the $\hat{y}$ direction, leading to a polar system with no inversion symmetry.
In this case, the induced orbital magnetization $\vec{M} \propto \vec{P} \times \vec{\mathcal J}$, which results in non-zero values of $M_z \propto P_y \mathcal{J}_x$. For nanoribbons with armchair borders, however, $P_y$ vanishes and so does the induced orbital magnetization, as our calculations confirm. 

\section{Conclusions} 

In summary, we have shown that nanoribbons with zigzag borders of $p_x$-$p_y$ band systems in a honeycomb lattice can exhibit fairly large orbital magnetization induced by an electric current flowing along the ribbon axis, when sublattice symmetry is broken. We have explored nanoribbons extracted from 2D systems in two distinct phases. In the first one, the 2D system is an ordinary insulator at the neutrality point and the ribbon exhibits no induced magnetization for Fermi-energies within the energy-band gap. Nevertheless, it shows significant values of OME outside this energy range, even in the absence of spin-orbit coupling. In the second case, the 2D system is a topological insulator that has conducting edge states crossing the energy band gap in the nanoribbon geometry. In this situation, the induced orbital magnetization is fairly large for energies within the 2D bulk gap and also attains high values of opposite sign just beyond the gap boundaries. In both cases, the induced orbital magnetization is oriented perpendicularly to the nanoribbon plane. Our results indicate that it may be controlled by a gate voltage, a basic requirement for device applications.

\begin{acknowledgments}
	We acknowledge CNPq/Brazil, CAPES/Brazil, FAPERJ/Brazil and INCT Nanocarbono for financial support.  TGR acknowledges funding from Fundação para a Ciência e a Tecnologia and Instituto de Telecomunicações - grant number UID/50008/2020 in the framework of the project Sym-Break. She thankfully acknowledges the computer resources at MareNostrum and the technical support provided by Barcelona Supercomputing Center (FI-2020-2-0033). LMC is supported by Project MECHANIC (PCI2018-093120) funded by the Ministerio de Ciencia, Innovación y Universidades. ICN2 is funded by the CERCA Programme/Generalitat de Catalunya and supported by the Severo Ochoa Centres of Excellence program, funded by the Spanish Research Agency (grant number SEV-2017-0706). FSMG gratefully acknowledge the computing time granted through JARA on the supercomputer JURECA \cite{jureca} at Forschungszentrum Jülich. Helpful conversations with Prof. P. Venezuela are gratefully acknowledged.
\end{acknowledgments}

%%%%%%%%%%%%%--APPendix--%%%%%%%%%%%%%%%%%%
\appendix

\section{Tight-binding model for zigzag nanoribbon}
\label{appA}
In this Appendix, we present details of the $p_x$-$p_y$ tight-binding model [Eq.~\eqref{Hpxpy}] used to compute the spin and OAM responses to an external electric field in ZZ nanoribbons.

\subsection{Kinetic Term}
The hopping amplitudes between orbitals on different sites $i$ and $j$ are computed using the Slater-Koster method~\cite{Slater-Koster}. They are written in terms of the direction cosines connecting sites $i$ and $j$, $n_{\mu}(i,j)$ ($\mu=x,y,z$), and of the two centers integrals, $V_{pp\sigma}$ and $V_{pp\pi}$,
\begin{eqnarray}
t_{\mu,\mu}(i,j)&=&n^2_{\mu}(i,j)V_{pp\sigma}+(1-n^2_{\mu}(i,j))V_{pp\pi}, \label{talphaalpha} \\
t_{\mu,\nu}(i,j)&=&-n_{\mu}(i,j)n_{\nu}(i,j)(V_{pp\pi}-V_{pp\sigma}), \label{talphabeta}
\end{eqnarray}
We restrict the hopping of our model to nearest neighbors sites, which have vectors, on the coordinate system of Fig. \ref{fig:figA2}(a), pointing along $\vec{e}_1=(a_0/\sqrt{3})(0,1)$, $\vec{e}_2=(a_0/\sqrt{3})(-\sqrt{3}/2,-1/2)$, $\vec{e}_3=(a_0/\sqrt{3})(\sqrt{3}/2,-1/2)$. 
Here, $a_0$ is the lattice constant. 
We now define the Slater-Koster matrix of hoppings connecting a site $i$ to its $m$-th neighbor ($m=1,2,3$) as
\begin{eqnarray}
t_m(k)=\sum_{\mu,\nu=x,y} \big|p_{\mu}\big> t_{\mu,\nu}(m)\exp (ik \ \vec{e}_m \cdot \hat{x}) \big<p_{\nu}\big|
\end{eqnarray}
where $t_{\mu,\nu}(m)$ are given by Eqs (\ref{talphaalpha}) and (\ref{talphabeta}). 
In the unit cell of the ZZ nanoribbon, there are two inequivalent sites; one at sublattice A and the other at sublattice B. The Hilbert space of the ZZ nanoribbon is spanned by $\mathcal{H}^{zz}= \{ \big|A,B\big>\otimes \big|p_x, p_y\big> \otimes \big|\ell \big> \} \otimes \big| k\big>$, where $\big|\ell \big>$ is the quantum number associated with the line of the ribbon. The hopping integral $t_{2,3}$ occurs within the same line $\ell$. We define the matrix of hoppings in the same line, in the basis $\big|A,B\big>$, as,
\begin{eqnarray}
 h_{l0}(k)=\begin{bmatrix}
   0 & t_2(k)+t_3(k) \\
    (t_2(k)+t_3(k))^{\dagger} & 0 
\end{bmatrix};
\end{eqnarray}
The matrices of hoppings that switch the lines are,
\begin{eqnarray}
 h_{l+}(k)=\begin{bmatrix}
   0 & t_1(k)\\
    0 & 0 
\end{bmatrix}; 
h_{l-}(k)=\begin{bmatrix}
   0 & 0\\
   t^{\dagger}_1(k) & 0 
\end{bmatrix}.
\end{eqnarray}
Using these matrices, we built the kinetic term of Hamiltonian of Eq.~\eqref{Hpxpy} in the reciprocal space for ZZ nanoribbon,
\begin{eqnarray}
H_{kin}(k)=\begin{bmatrix}
    h_{l0}(k) & h_{l-}(k) & 0 & \dots  & 0 \\
    h_{l+}(k) & h_{l0}(k)& h_{l-}(k) & \dots  & 0 \\
    0 &h_{l+}(k) &  h_{l0}(k) & \dots  & 0 \\
    \vdots & \vdots & \vdots & \ddots & \vdots \\
    0 & 0 & 0 & \dots  & h_{l0}(k)
\end{bmatrix}.
\label{Hzz}
\end{eqnarray}
For a ribbon with $N$ lines, the matrix of Eq.~\eqref{Hzz} have a tri-diagonal block form with $N$ diagonal blocks $h_{l0}(k)$.

\subsection{SOC and Sublattice Potential}
The component $L^z$ of the orbital angular momentum operator at atomic-approximation \cite{GoXueLinearResponse, AtomCentred_Modernteo_2, Orbital-Magnetism-Go} in the basis $\{ \big|p_x\big>, \big|p_y\big>\}$ is written as,
\begin{eqnarray}
L^z=\begin{bmatrix}
   0 & -i\\
    i & 0 
\end{bmatrix}.
\end{eqnarray}
The spin-orbit coupling term is written as,
\begin{eqnarray}
H_{\text{SOC}}=\lambda_{I} S^zL^z \mathbb{1}_{\sigma} \mathbb{1}_{\ell},
\label{HSOC}
\end{eqnarray}
where, $S^z=\text{diag}(1,-1)$ is the Pauli matrix related to electron spin, and $\mathbb{1}_{\sigma}$ ($\mathbb{1}_{\ell}$) is the identity operator in sublattice (line) degree of freedom. The sublattice potential term is defined as,
\begin{eqnarray}
H_{\text{AB}}=V_{\text{AB}} \mathbb{1}_{S}\mathbb{1}_{L} \sigma_z \mathbb{1}_{\ell},
\label{HVAB}
\end{eqnarray}
where, now, $\sigma_z=\text{diag}(1,-1)$ is the Pauli matrix related to sublattice degree of freedom and $\mathbb{1}_{S}$($\mathbb{1}_{L}$) is the identity operator in the spin (orbital) space.

%%%%%%%%%%%%--Bibliography--%%%%%%%%%%%%%%%%%

\bibliographystyle{apsrev}
%\bibliography{bibliografia}

\end{document}